\documentclass[sigconf, nonacm]{acmart}
\usepackage{enumitem}
\usepackage{appendix}
\usepackage{listings}
\usepackage{xcolor}
\usepackage{titlesec}
\usepackage[utf8]{inputenc}   
\usepackage[T1]{fontenc}      
\usepackage{geometry}         
\usepackage{tcolorbox}        
\AtBeginDocument{%
  }

\setcopyright{acmlicensed}
\copyrightyear{2018}
\acmYear{2018}
\acmDOI{XXXXXXX.XXXXXXX}
\acmConference[Conference acronym 'XX]{Make sure to enter the correct
  conference title from your rights confirmation email}{June 03--05,
  2018}{Woodstock, NY}
\acmISBN{978-1-4503-XXXX-X/2018/06}

\begin{document}




\title{SimuPanel: A Novel Immersive Multi-Agent System to Simulate Interactive Expert Panel Discussion}

\author{Xiangyang He}
\authornote{Both authors contributed equally to this research.}
\email{xiangyanghe@hkust-gz.edu.cn}
\affiliation{%
  \institution{The Hong Kong University of Science and Technology (Guangzhou)}
  \city{Guangzhou}
  \country{China}
}

\author{Jiale Li}
\authornotemark[1]
\email{jli022@connect.hkust-gz.edu.cn}
\affiliation{%
  \institution{The Hong Kong University of Science and Technology (Guangzhou)}
  \city{Guangzhou}
  \country{China}
}

\author{Jiahao Chen}
\email{2251646@tongji.edu.cn}
\affiliation{%
  \institution{Tongji University}
  \city{Shanghai}
  \country{China}
}

\author{Yang Yang}
\email{yyiot@hkust-gz.edu.cn}
\affiliation{%
  \institution{The Hong Kong University of Science and Technology (Guangzhou)}
  \city{Guangzhou}
  \country{China}
}

\author{Mingming Fan}
\email{mingmingfan@ust.hk}
\authornote{Corresponding author}
\affiliation{%
  \institution{The Hong Kong University of Science and Technology (Guangzhou)}
  \city{Guangzhou}
  \country{China}
}


\begin{abstract}
Panel discussion allows the audience to learn different perspectives through interactive discussions among experts moderated by a host and a Q\&A session with the audience. Despite its benefits, panel discussion in the real world is inaccessible to many who do not have the privilege to participate due to geographical, financial, and time constraints. We present \textit{SimuPanel}, which simulates panel discussions among academic experts through LLM-based multi-agent interaction. It enables users to define topics of interest for the panel, observe the expert discussion, engage in Q\&A, and take notes. \textit{SimuPanel} employs a host-expert architecture where each panel member is simulated by an agent with specialized expertise, and the panel is visualized in an immersive 3D environment to enhance engagement. Traditional dialogue generation struggles to capture the depth and interactivity of real-world panel discussions. To address this limitation, we propose a novel multi-agent interaction framework that simulates authentic panel dynamics by modeling reasoning strategies and personas of experts grounded in multimedia sources. This framework enables agents to dynamically recall and contribute to the discussion based on past experiences from diverse perspectives. Our technical evaluation and the user study with university students show that \textit{SimuPanel} was able to simulate more in-depth discussions and engage participants to interact with and reflect on the discussions. As a first step in this direction, we offer design implications for future avenues to improve and harness the power of panel discussion for multimedia learning.
\end{abstract}


\begin{teaserfigure}
  \includegraphics[width=\textwidth]{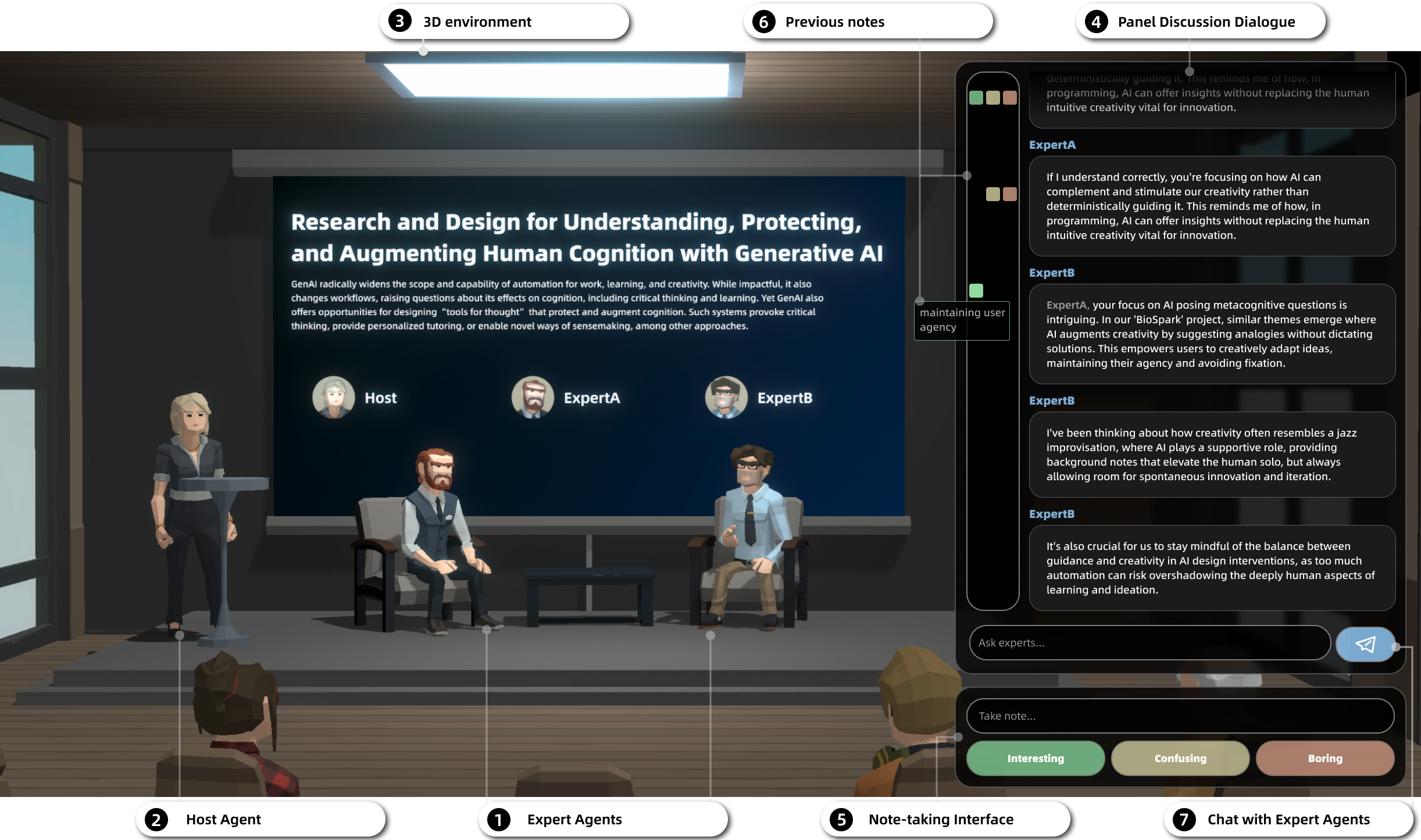}
  \caption{\textit{SimuPanel} Interface Overview. The system simulates academic panel discussions with interactive expert agents. Users can view the topic poster, follow the live multimodal dialogue, and access a scrollable discussion history. Note-taking is supported via keyboard shortcuts with color-coded labels for quick categorization.}
  \Description{User Interface.}
  \label{fig:teaser}
\end{teaserfigure}


\maketitle

\section{Introduction}
Academic panel discussions at conferences offer audiences a unique opportunity to learn from experts who exchange viewpoints, engage in real-time interaction, and collaboratively explore complex topics through both prepared remarks and spontaneous dialogue~\cite{doumont2014english}. These panel formats bring together scholars who challenge assumptions, build on each other's ideas, and allow audience members to observe the dynamic process of knowledge construction. Such discussions have proven effective in clarifying differing perspectives for attendees and encouraging high levels of participation~\cite{gallo2020grant}. Their value has persisted over the years across a range of disciplines, from finance~\cite{summers1991panel} and environmental science~\cite{foken2011results} to social science~\cite{wang2017longitudinal}.

Panel discussions also reveal something rarely accessible in traditional instructional formats such as textbooks or lectures: the way experts think, debate, and refine ideas. This connects closely to cognitive apprenticeship theory, which emphasizes the importance of making thinking visible. As Collins et al.~\cite{collins2006cognitive} note, "the processes of thinking are often invisible to both the students and the teacher." In contrast, academic panels make these mental processes explicit, through structured argumentation, active questioning, and collaborative knowledge construction. These interactive panels, where experts from similar or diverse disciplines respond to each other at the moment, create a rich learning environment that models advanced academic conversation. 

Despite their educational value, access to high-quality academic panel discussions remains limited. These events typically occur at specialized conferences or during interdisciplinary workshops, settings that are geographically restricted, temporally finite, and often accessible only to those within specific academic communities. Many students and researchers, particularly those from institutions with fewer resources or in regions with limited academic networking opportunities, miss these valuable learning experiences in terms of registration fees and travel barriers~\cite{wu2022virtual}. Available alternatives, such as published discussions that lack interactivity or selective video clips from conferences, fail to capture the dynamic, dialogical nature of live panel discussions.

This limited access to real-world panel discussions presents one unique opportunity, especially given recent advances in Large Language Models (LLMs) and multi-agent systems~\cite{park2023generative}. These technologies have demonstrated promising capabilities in generating coherent conversations~\cite{sekulic2024analysing}, reasoning through complex topics~\cite{guo2025deepseek}, and even simulating interactions between distinct personas~\cite{chenpersona}. Building on these advancements, an opportunity emerges: \textit{what if we could simulate academic panel discussions that bring together the distributed insights of multiple experts?} Such a system would not only consolidate knowledge from disparate sources from papers to talks but also model the dynamic exchange of ideas that occurs when experts engage with each other's perspectives. However, our preliminary experiments revealed that basic prompt-based LLM agents with real-world grounded information cannot generate sufficiently in-depth discussions (as demonstrated in Figure~\ref{fig:reasoning_strategies}), highlighting the need for better approaches.

To address these challenges, we introduce \textit{SimuPanel}, a novel multimedia learning system that help learners access to academic panel discussions through multi-agent simulation. Unlike existing approaches that rely on simple prompt-based generation, \textit{SimuPanel} addresses the core challenges of authentic expert representation, coherent multi-agent dialogue, and engaging educational interfaces. The system bridges the gap between the rich pedagogical value of live academic panels and the accessibility needs of learning communities, offering a scalable solution. 

Our key contributions are threefold:

\begin{itemize}[topsep=0pt, itemsep=2pt, parsep=0pt, leftmargin=1.5em]
    \item \textbf{A novel multi-agent framework for academic panel simulation} that combines two-layer persona construction, modular reasoning processes, and structured discussion flow management. Our ablation study demonstrates that this integrated approach significantly outperforms simpler baselines across six evaluation dimensions.
    \item \textbf{An immersive multimedia learning interface} featuring 3D expert avatars with with multimodal representations, interactive color-coded note-taking, and follow-up question capabilities that enhance user engagement and support active learning during simulated panel discussions.
    \item \textbf{Comprehensive system evaluation and design insights} through LLM-based comparative analysis and a user study with 10 graduate students, revealing that \textit{SimuPanel} effectively supports learning and inspiration while identifying key design considerations for future educational simulation systems.
\end{itemize}

\section{Related Work}
\subsection{Multimedia Learning System}

Multimedia learning systems encompass a variety of platforms designed to enhance knowledge construction by integrating visual and auditory information, spanning static formats to highly interactive environments such as animations, videos, and immersive simulations~\cite{mayer2002multimedia}. In recent years, these systems have significantly expanded access to educational content, leveraging technological advancements to address diverse learner preferences and instructional scenarios.

Early multimedia learning platforms, notably MOOCs, initially sought to democratize higher education by providing widespread access to high-quality instructional materials~\cite{reich2019mooc}. However, substantial research has revealed critical limitations in achieving their effect, including persistently low learner retention and completion rates, high early-course attrition, and an imbalance favoring affluent learners from developed regions~\cite{reich2019mooc}. Such platforms struggled particularly with sustaining learner engagement over time, highlighting challenges in maintaining consistent interest and participation beyond initial enrollment. In response to these challenges, newer multimedia systems have increasingly adopted immersive technologies. These advanced platforms demonstrate strong potential to address declining learner engagement by creating interactive and realistic learning experiences. Recent empirical studies indicate that the immersive environment can enhance cognitive, affective, and behavioral engagement, sustaining learner interest by simulating authentic contexts and interactive scenarios~\cite{lin2024impact}. For instance, immersive virtual classrooms can closely replicate face-to-face interactions and promote active participation~\cite{florez2023evaluating}.

Parallel to these technological advancements, the emergence of generative AI, particularly large language models (LLMs), has enabled the dynamic generation of conversational learning content. For instance, Google’s NotebookLM simulates podcast-style discussions derived from user-uploaded documents. However, Huffman and Hutson~\cite{huffman2024enhancing} found that current implementations offer limited control over content scope and conversational direction, reducing their alignment with structured learning objectives. These limitations highlight the need for systems that support both interactive discussion and user-guided personalization.

\subsection{Role-Playing Language Agents}

Recent advances in large language models (LLMs) have enabled the emergence of Role-Playing Language Agents (RPLAs)—virtual agents capable of simulating human-like reasoning, behavior, and interaction across a variety of contexts~\cite{shanahan2023role, chen2024persona}. RPLAs are increasingly adopted in applications ranging from social behavior simulation~\cite{park2023generative, kaiya2023lyfe} and collaborative problem-solving~\cite{du2023improving, qian2023communicative} to intelligent game copilots~\cite{wang2023voyager, xu2023exploring}. In the context of educational discussion, these agents offer a powerful foundation for modeling academic conversations, where reasoning, dialogue, and knowledge construction unfold dynamically.

A foundational component of RPLA design is the construction of persona—a structured representation of an agent's identity, expertise, and behavioral tendencies. Personas enable agents to exhibit coherent, individualized behaviors that reflect real-world academic profiles~\cite{dong2024self, park2022social}. Recent frameworks categorize personas into demographic, character, and individualized types~\cite{chen2024persona}, with the Individualized Persona being especially relevant for simulating domain experts. This approach leverages public profiles to capture the intellectual stance, interests, and discussion patterns of specific individuals~\cite{shao2023character}. To support adaptive and context-sensitive behavior, persona creation often integrates pre-defined structures, model-generated traits, and data-derived insights~\cite{guo2024large, weiss2024rethinking}.

However, simulating expert discussion requires more than static persona representation—it necessitates sophisticated reasoning capabilities that enable agents to engage in back-and-forth dialogue, respond to opposing views, and refine ideas collaboratively. The reasoning processes in RPLAs typically involve multi-stage workflows composed of memory, reflection, planning, and deliberation~\cite{topsakal2023creating}. These workflows are supported by LLM inference capabilities and prompting strategies, such as Chain-of-Thought~\cite{wei2022chain} and Tree-of-Thought~\cite{long2023large}, which have been shown to improve the ability of LLMs to handle complex, multi-step reasoning~\cite{sprague2024cot}. Reinforcement learning-based strategies further enhance these models by training agents to pursue long-term reasoning goals aligned with conversational or instructional outcomes~\cite{guo2025deepseek}.

RPLAs have become central to multi-agent simulation systems, where agents interact under structured communication paradigms, such as cooperative, competitive, or debate-based models—to emulate human discussion~\cite{guo2024large}. Debate paradigms are particularly well-suited for simulating academic panels, as they promote critical evaluation, diverse perspectives, and collaborative knowledge construction~\cite{du2023improving}. Leveraging RPLAs in this context enables the modeling of individualized expertise, contextual reasoning, and interactive dialogue. Unlike scripted or Q\&A-based systems, RPLAs can reproduce the fluid, adaptive nature of real scholarly discussions, forming a robust foundation for systems like \textit{SimuPanel}.

\section{Design of \textit{SimuPanel}}

Building on prior work in multimedia learning systems and LLM-based agent simulation, we designed \textit{SimuPanel} to simulate academic panel discussions that promote learner engagement, make expert reasoning visible, and support personalized topic exploration. This section outlines three primary design considerations (\textbf{DC}) that guided our approach:

\textbf{DC1: Authentic Expert Representation.} A core design principle in \textit{SimuPanel} is the representation of experts as distinct academic personas with coherent viewpoints and reasoning styles. Rather than treating experts as generic knowledge repositories, we envisioned agents that would embody the interpretive positions and intellectual approaches characteristic of real scholars~\cite{chen2024persona, guo2024large}. This design choice was informed by research showing that learners benefit from observing how experts interpret evidence, form arguments, and respond to alternative perspectives~\cite{christakopoulou2024agents}. By exposing these interpretive processes, \textit{SimuPanel} aims to provide learners with insights into disciplinary thinking that typically remains implicit in traditional educational materials. Drawing from Clark et al.'s~\cite{clark2019makes} research on high-quality human conversation, we prioritized three conversational attributes in our design: mutual understanding between experts, trust-building through acknowledgment of others' contributions, and active listenership demonstrated through responsive engagement with others. These principles guided our approach to creating expert agents capable of engaging in naturalistic discourse.

\textbf{DC2: Coherent Multi-Agent Discussion Modeling.} Our second design consideration focused on creating a framework for coordinated, purposeful dialogue between multiple expert agents. Rather than presenting isolated viewpoints, we sought to model the dynamic interplay of ideas characteristic of scholarly panel discussions. This approach was informed by research on debate-style communication paradigms that support critical evaluation and collaborative knowledge construction~\cite{du2023improving, guo2024large}. We conceptualized panel discussions as progressing through distinct stages, opening remarks, main discussion with moderated exchanges, and closing synthesis with audience engagement, each with specific communicative goals. This staged approach was designed to provide structural scaffolding for learners while maintaining the organic flow of ideas essential to authentic academic discourse. By modeling these interaction patterns, \textit{SimuPanel} aims to make visible not just what experts know, but how they engage with and build upon each other's ideas.

\textbf{DC3: Educational Engagement.}  
Our third design consideration focuses on supporting learning engagement.Our design emphasizes immersive and interactive features that help learners stay cognitively connected to the discussion. Drawing on insights from multimedia learning~\cite{florez2023evaluating, lin2024impact}, we propose a 3D environment that provides visual and temporal cues to scaffold comprehension of conversational flow and speaker dynamics. Additionally, our approach integrates research on note-taking and adaptive learning~\cite{piolat2005cognitive, el2021adaptive, gligorea2023adaptive}, emphasizing features that reduce cognitive overload while promoting active participation and learner agency. These design principles collectively aim to create an experience that is both intellectually rigorous and broadly accessible, bringing the pedagogical value of academic panels to a wider and more diverse audience.

\section{Implementation of \textit{SimuPanel}}

\subsection{Agent Framework (DC1)}
Our framework implements a cognitive model for expert panelists in academic discussions, structured around a two-layer persona and modular reasoning processes. Each agent is instantiated with a distinct academic identity and knowledge base. Formally, each expert agent $A = (P, K, R, U)$ consists of a persona $P$, knowledge base $K$, reasoning process $R$, and utterance generation mechanism $U$. The agent's persona comprises two information layers: a low-level layer storing domain knowledge and research experience, and a high-level layer capturing research interests and beliefs.

\begin{figure*}[h] 
\centering
\includegraphics[width= 1\textwidth]{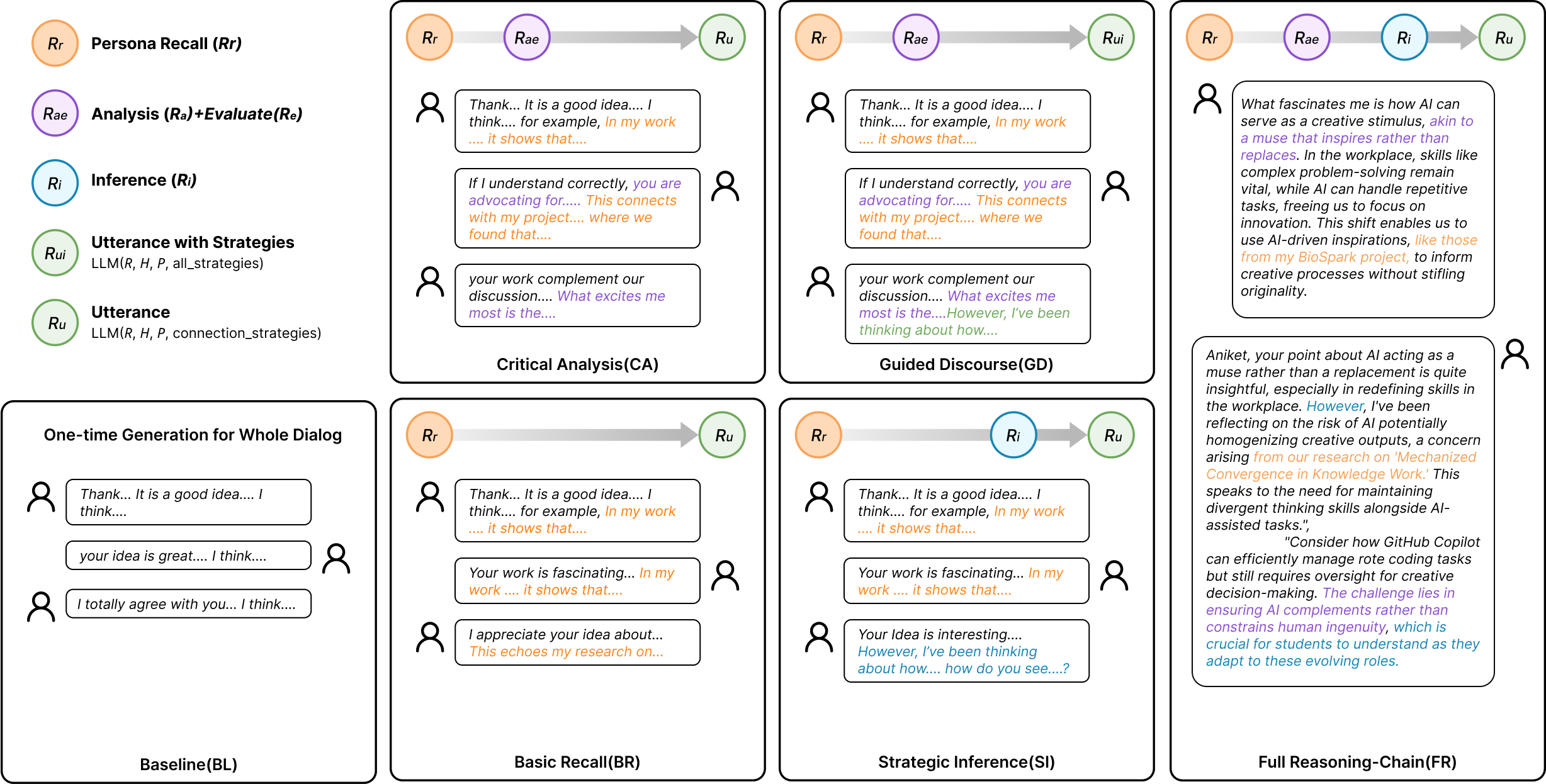}
\caption{The figure illustrates the reasoning workflows of the six strategies evaluated in our ablation study. We present structural descriptions of the generated content for BL, BR, SI, CA, and GD, as well as a complete dialogue excerpt for FR. Different colored text highlights the specific functions of the corresponding modules within the dialogue: for example, orange indicates the Recall module, which drives the inclusion of relevant knowledge references; purple denotes the Analysis module, which leads to more detailed viewpoints and clarification; and blue represents the Inference module, which introduces more diverse dialogue styles, such as questions and challenges. Green represents the approach where all available dialogue strategies are provided during the utterance generation phase. This approach achieves effects similar to those of the Inference module, as the agent adopts questioning and challenging dialogue strategies, but with a lower frequency.}
\label{fig:reasoning_strategies}
\end{figure*}

We construct this two-layer persona using both data-driven and model-generated approaches. For the low-level layer, we incorporate an automation script to collect relevant homepages, academic articles, and public talks from open-access platforms (e.g., Google Scholar, YouTube), storing titles, abstracts, and segmented content chunks for vectorized retrieval. The high-level layer is constructed by summarizing the low-level data using GPT-based models with prompts to identify pre-established stances relevant to the panel topic. For knowledge retrieval, we implement a retrieval augmented generation (RAG) approach where relevant knowledge chunks are fetched based on semantic similarity to the current discussion topic. These chunks undergo a GPT-4-based filtering process to select information most pertinent to downstream tasks. The reasoning process $R$ is implemented as a modular pipeline with components that can be combined in various ways shown in Figure~\ref{fig:reasoning_strategies}.

\begin{equation}
R(q, K, H) = R_{inference} \circ R_{evaluate} \circ R_{analysis} \circ R_{recall}(q, K, H)
\end{equation}

Where $H$ represents dialogue history, and each component performs a specific cognitive function. The Recall module retrieves relevant knowledge. The Analysis module synthesizes research perspectives. The Evaluate module enables critical reflection on retrieved content. The Inference module handles strategic discussion selection, optimizing from strategies like questioning, answering, scholarly agreement, constructive critique, and synthesis.

Additionally, we analyze the behaviors of participants in real-world panel meetings to construct a strategy module tailored to the panel meeting scenario. This enables agents to exhibit interaction behaviors that more closely align with those observed in real-world academic discussions, as emphasized in recent agent-based simulation system designs\cite{wang2025limits}. Within the Inference module, the discussion strategy selection optimizes 
from a set of strategies 
\[
S = \left\{
\begin{array}{l}
\text{question},\ \text{answer},\ \text{scholarly\_agreement},\\
\text{constructive\_critique},\ \text{synthesis}
\end{array}
\right\},
\]
selecting the strategy \( s^* \) that maximizes both educational value and 
alignment with the agent's beliefs:

\begin{equation}
s^* = \arg\max_{s \in S} \{\text{educational\_value}(s) + \text{belief\_alignment}(s, P.\text{beliefs})\}
\end{equation}

Where $\text{educational\_value}(s)$ represents the potential instructional benefit of a strategy, and $\text{belief\_alignment}(s, P.\text{beliefs})$ measures how well the strategy aligns with the agent's established viewpoints.

The utterance generation $U$ then synthesizes all reasoning components into a coherent verbal contribution:

\begin{equation}
U(R, H, P) = \text{LLM}(R, H, P, \text{connection\_strategies})
\end{equation}

Where $\text{LLM}$ represents the large language model that generates the final response based on the reasoning output $R$, dialogue history $H$, persona $P$, and connection strategies that include elements of mutual understanding, trust-building, and active listening to create authentic academic discussion.

Figure~\ref{fig:reasoning_strategies} illustrates the reasoning processes of different reasoning strategies and provides a feature analysis of the panel discussion samples generated by each strategy.

\subsection{Multi-agent Interaction Mechanism (DC2)}
\textit{SimuPanel} implements a structured multi-agent interaction framework designed to emulate authentic academic panel discussions, where a host agent coordinates the dialogue among expert agents to ensure topical relevance, balanced participation, and coherent conversational progression.

The host agent serves as a virtual moderator responsible for managing turn-taking, orchestrating topic transitions, and maintaining discussion momentum. Its behavior is governed by an internal policy defined as $\mathcal{H} = \{\mathcal{S}, \mathcal{T}, \mathcal{A}, \mathcal{P}\}$, where $\mathcal{S}$ denotes the current discussion stage (e.g., opening, discussing, converging, or end), $\mathcal{T}$ represents the topic set, $\mathcal{A}$ contains agenda criteria for each stage, and $\mathcal{P}$ defines the progression rules between stages.

An internal state machine governs the panel's stage progression. Each stage is associated with distinct communicative goals: establishing a foundation in the opening, exploring depth during discussion, synthesizing insights in convergence, and summarizing key takeaways in the closing stage. Transitions between stages are driven by real-time evaluation of expert interactions against stage-specific criteria.

To maintain coherent discussion, \textit{SimuPanel} employs a recurrent cycle of evaluation and intervention. At each turn $t$, the host assesses the ongoing conversation and selects a moderation action defined as $\mathcal{D}_t = f_{LLM}(H_t, S_t, C_t, B_t)$, where $\mathcal{D}_t$ is a discrete decision variable with possible values CONTINUE, TRANSITION, or REDIRECT. The inputs include $H_t$, the recent dialogue history; $S_t$, the current discussion stage; $C_t$, the topic under discussion; and $B_t$, a contextual bias term that adjusts intervention sensitivity based on factors such as prolonged exchanges without moderation or unresolved prompts. The decision $\mathcal{D}_t$ determines whether to allow the discussion to continue, transition to a new topic or stage, or redirect the current dialogue flow.

To regulate pacing and cognitive load, the system applies bounded turn management by enforcing minimum and maximum exchange thresholds between host interventions ($\tau_{\text{min}} = 3$, $\tau_{\text{max}} = 6$). This ensures sufficient dialogue depth while triggering timely moderation to maintain coherence and learner engagement. When intervention is warranted, the host generates contextually appropriate responses that acknowledge previous contributions, synthesize key points, and guide the discussion forward.

\subsection{Interactive Interface (DC3)}

\textit{SimuPanel} provides an interactive interface shown in Figure~\ref{fig:teaser} that combines immersive visuals with lightweight learning tools. Drawing inspiration from real-world academic panels, we developed a stylized 3D environment in Unity where avatar agents deliver speech by Text-To-Speech (TTS)~\cite{anastassiou2024seed} accompanied by synchronized gestures and movements. These nonverbal cues help users follow the flow of conversation, speaker turns, and rhetorical emphasis, enhancing comprehension and presence. To support efficient learning, the system includes a color-coded note-taking feature that allows users to annotate key ideas as the panel unfolds. These notes act as cognitive anchors, helping learners revisit important moments and reflect on the discussion afterward.

\textit{SimuPanel} also introduces a \textit{student audience agent}, a virtual participant that formulates a single, thoughtful question after the panel. Based on the full dialogue and grounded in a real-world example, the question is directed to a relevant expert, encouraging deeper reflection and modeling critical inquiry.

Additionally, users can pose their own follow-up questions to expert agents after the session. Responses are generated in context, enabling learners to clarify concepts or explore related topics beyond the original discussion.

\section{Technical Evaluation}
\subsection{Dataset Preparation}
To evaluate \textit{SimuPanel}’s capacity for generating meaningful, learning-oriented discussions, we collected five recent panel information from three top-tier Human-Computer Interaction conferences, including CHI\footnote{\href{https://chi2025.acm.org/for-authors/workshops/accepted-workshops/} Human Factors in Computing Systems Workshops in 2025.}, CSCW\footnote{\href{https://programs.sigchi.org/cscw/2024/program/} Computer-Supported Cooperative Work and Social Computing Panels in 2024.}, and UIST\footnote{\href{https://programs.sigchi.org/uist/2024/program/}Symposium on User Interface Software and Technology Workshops in 2024.}. These panels address timely and high-impact topics, including social technologies, sociotechnical ethics, affective computing, human cognition, and intelligent materials, offering a testbed for assessing \textit{SimuPanel}'s ability to model expert reasoning, debate, and reflective dialogue. Our selection criteria include (1) relevance to open-ended and critical questions, (2) potential to inspire deeper inquiry among university students in the HCI community, and (3) comprehensive materials, including panelist names, detailed abstracts, and discussion questions. For simulation, we selected three participants per panel to form the host–expert configuration.

Our experiment involved generating six distinct dialogues based on ChatGPT-4o for ablation study under each of the five topics. Furthermore, to account for the inherent stochasticity of LLM generation, we generated each dialogue twice under identical topic and reasoning conditions.

\subsection{LLM-based Evaluation}
To assess the quality of the generated panel dialogues, we adopted an LLM-based evaluation approach using LLMs as judges. Inspired by recent work such as ResearchAgent~\cite{baek2024researchagent} and AI-Scientist~\cite{lu2024ai}, which demonstrated the effectiveness of LLMs for dialogue assessment, we followed a similar methodology from Li et al.~\cite{li2024chain}, which provided a viable approach to the problem of open-ended content quality comparison. They showed that LLM-based judgments align closely with human evaluations, using a Round-Robin tournament to compute ELO scores. Our evaluation criteria were adapted from Fu et al.~\cite{fu2023gptscore} and included six dimensions: specificity, relevance, flexibility, coherence, informativeness, and depth of analysis. We used the DeepSeek-R1 model~\cite{guo2025deepseek} for its ability to produce detailed and stepwise analysis. 

During the evaluation process, the model provided comparative judgments across all six metrics, assigning winner-number outcomes to indicate which dialogue performed better or resulted in a close stance. These numerical count assessments were accompanied by detailed justifications that offered qualitative insights into each dialogue's strengths and weaknesses. After computing the ELO score for each reasoning process across the five topics, we calculated the average score. To mitigate potential biases and ensure robustness, we evaluated each pair of dialogues in both possible orders.  We then computed a final average across our two-generation runs to produce our definitive results.

\subsection{Ablation Study Results}

To understand the impact of different reasoning components on dialogue quality, we conducted an ablation study. Figure~\ref{fig:reasoning_strategies} illustrates these strategies, which progressively incorporate different cognitive processes. Our ablation study examined the following configurations:

\begin{itemize}[topsep=0pt, itemsep=2pt, parsep=0pt, leftmargin=1.5em]
    \item \textbf{Baseline (BL)}: Uses a simple one-shot generation without any agentic workflow for the whole dialog.
    
    \item \textbf{Basic Recall (BR)}: Implements a minimal reasoning chain with only knowledge recall before generating responses.
    
    \item \textbf{Critical Analysis (CA)}: Extends the BR approach by adding analysis and evaluation steps to critically assess the discussion.
    
    \item \textbf{Guided Discourse (GD)}: Augments The impact of critical analysis results on utterance output with explicit discourse guidance, offering agents multiple communication strategies to choose from.

    \item \textbf{Strategic Inference (SI)}: Incorporates an inference step after knowledge recall, allowing agents to select and justify discourse strategies based on their expertise.
    
    \item \textbf{Full Reasoning-chain (FR)}: Combines all components—knowledge recall, analysis, evaluation, and Incorporates an inference step before utterance, allowing agents to focus on the selected strategy during utterance generation.
\end{itemize}


As shown in Table~\ref{tab:evaluation_results}, the \textbf{FR} strategy outperformed all other approaches (38.17 total points). Comparing FR with SI (38.17 vs. 22.49) reveals that \textbf{analytical processing} substantially enhances the effectiveness of strategic inference, with particularly notable improvements in Specificity (+3.0), Informativeness (+3.5), and Depth of Analysis (+3.5).

The \textbf{GD} approach ranked second (31.00), demonstrating that \textbf{explicit discourse guidance} significantly improves performance over BR (+11.0 total points). This improvement was most pronounced in Flexibility (+2.16) and Specificity (+2.34), suggesting that structured communication strategies enhance adaptability while maintaining focus, even without analytical components.

The \textbf{CA} strategy placed third (25.00), outperforming BR by 5.0 points. This improvement primarily came from enhanced Depth of Analysis (+0.83) and Specificity (+1.0), confirming that \textbf{analytical processing} contributes to more substantive content. However, the modest gains compared to GD suggest that analytical processing alone is insufficient without strategic guidance.

Comparing \textbf{SI} (fourth, 22.49) with BR (fifth, 20.00) shows that \textbf{strategic inference} provides moderate benefits (+2.49 points), with the most significant improvements in Relevance (+2.0). However, the substantial performance gap between SI and FR indicates that strategic inference reaches its full potential only when combined with analytical processing.

The \textbf{BR} approach performed poorly (20.00), particularly in Relevance (1.33), highlighting that mere knowledge recall without analytical or strategic components leads to responses that often miss the contextual needs of the discussion. The \textbf{BL} performed substantially worse than all structured reasoning approaches (6.67).
\begin{table*}
    \centering
    \caption{Performance comparison of six reasoning strategies across evaluation criteria. Results demonstrate the superiority of integrated reasoning approaches (Full Reasoning-chain).}
    \label{tab:evaluation_results}
    \resizebox{\textwidth}{!}{%
    \begin{tabular}{lccccccccc}
        \hline
        \textbf{Group} & 
        \textbf{Specificity} & 
        \textbf{Relevance} & 
        \textbf{Flexibility} & 
        \textbf{Coherence} & 
        \textbf{Informativeness} & 
        \textbf{Depth of Analysis} & 
        \textbf{Avg} & 
        \textbf{Total} &
        \textbf{Rank} \\
        \hline
        Full Reasoning-chain        & \textbf{7.33} & \textbf{4.17} & 5.67 & \textbf{6.17} & \textbf{7.33} & \textbf{7.50} & \textbf{6.36} & \textbf{38.17} & \textbf{1} \\
        Guided Discourse         & 6.17 & 2.67 & \textbf{5.83} & 4.83 & 5.67 & 5.83 & 5.17 & 31.00 & 2 \\
        Critical Analysis        & 4.83 & 2.83 & 3.67 & 4.17 & 4.50 & 5.00 & 4.17 & 25.00 & 3 \\
        Strategic Inference         & 4.33 & 3.33 & 3.17 & 3.83 & 3.83 & 4.00 & 3.75 & 22.49 & 4 \\
        Basic Recall        & 3.83 & 1.33 & 3.67 & 3.00 & 4.00 & 4.17 & 3.33 & 20.00 & 5 \\
        Baseline & 1.50 & 1.17 & 0.67 & 1.33 & 1.00 & 1.00 & 1.11 & 6.67  & 6 \\
        \hline
    \end{tabular}    }
\end{table*}
\section{User Study}

\subsection{Study Objectives and Design}
We conducted a user study with 10 graduate HCI students (ages 22-29) to evaluate \textit{SimuPanel}'s learning effectiveness and learning experience. In this study, participants selected two topics from a set of five related to their academic background, with the topic selection criteria outlined in Section 5.1. They then engaged in corresponding simulated panel sessions. During each session, they could take notes to record key ideas and reflections. Afterward, they had the option to interact with expert agents to further explore topic-related questions. 

For learning effectiveness measurement, participants answered two open-ended questions related to the selected topic before and after each session, allowing us to assess whether their responses reflected new perspectives or inspiration. They also completed a nine-item, five-point Likert scale on self-perceived learning effectiveness, with follow-up questions to explain their ratings. For learning experience evaluation, participants completed the System Usability Scale (SUS)~\cite{brooke1996sus} after both sessions, followed by semi-structured interviews on specific system features.
                                                                                 
\subsection{User Study Results}
\begin{figure}[t]
  \centering
  \includegraphics[width=\linewidth]{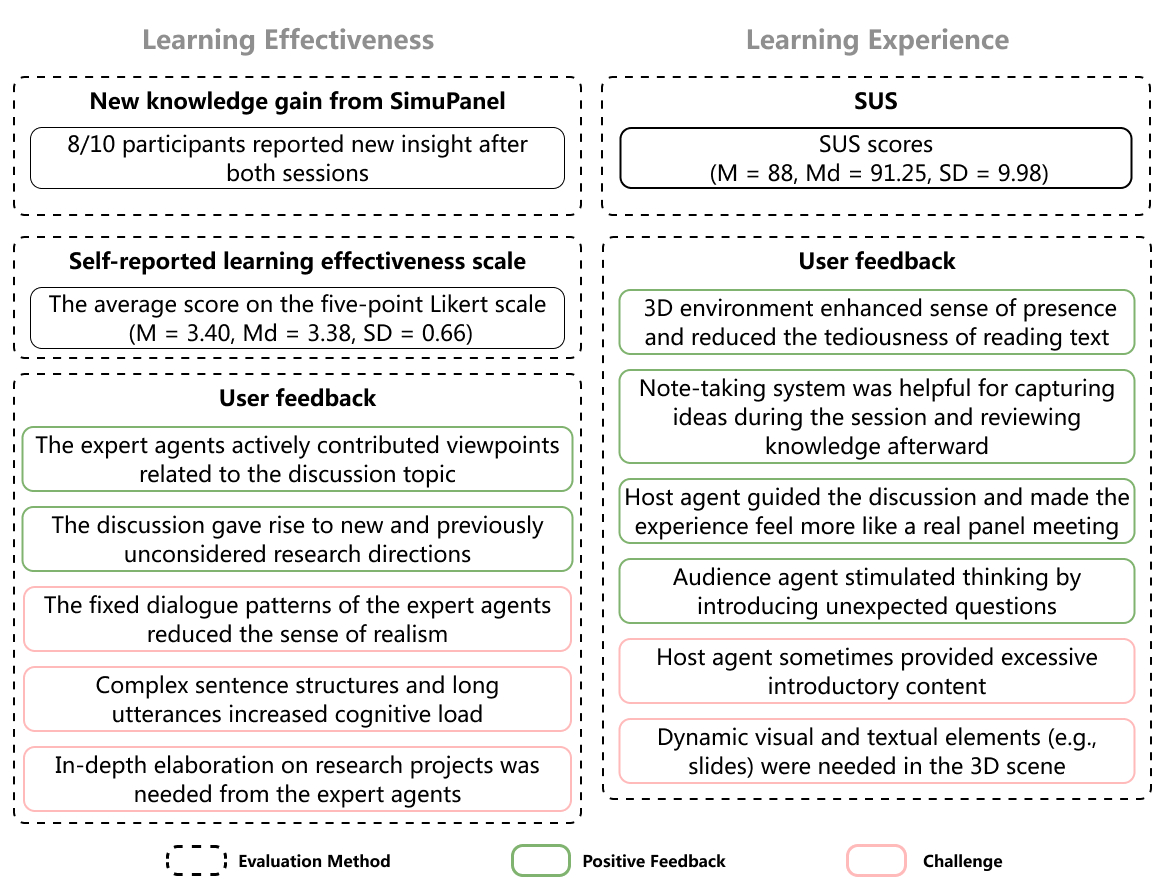}
  \caption{Summary of user study results evaluating the learning effectiveness and experience of \textit{SimuPanel}. Left: Effectiveness is assessed through knowledge gain, self-reported learning effectiveness scale, and participants’ explanations of ratings. Right: Learning experience is evaluated via SUS scores and qualitative feedback in semi-structured interviews.}
  \label{fig:userstudy}
  \end{figure}

Figure~\ref{fig:userstudy} summarizes our findings on learning effectiveness and user experience. This study provides a preliminary validation of the effectiveness of using \textit{SimuPanel} for learning, evaluates the overall user experience with the system, and identifies key design features that influence both learning effectiveness and user experience.

\subsubsection{Learning effectiveness.}
Most participants (8/10) demonstrated clear learning gains by sharing new insights after both \textit{SimuPanel} sessions, with the remaining two acquiring new knowledge in only one session. The average score on self-reported learning effectiveness was moderate (M = 3.40, Md = 3.38, SD = 0.66).

Participants who rated the system highly appreciated the coherent dialogue between expert agents and the exchange of viewpoints that offered research inspiration. As P1 noted, "\textit{The distinction between AI as a tool and as a partner offers a new perspective for understanding the collaborative relationship between humans and AI.}" P6 reflected that "\textit{The discussion highlighted bias and cross-cultural issues, which I had previously overlooked when implementing Emotion AI.}"

Those with lower ratings identified limitations in realism and depth. Several participants noted fixed dialogue patterns becoming apparent as conversations progressed. P5 found the dialogue challenging to follow due to complex sentence structures and lengthy utterances. Others desired more concrete explanations about research methodologies, with P5 noting that "\textit{The discussion emphasized storytelling projects related to social technologies, but did not elaborate on the specific design methodologies behind them.}"

\subsubsection{Learning experience.}
\textit{SimuPanel} received high usability scores (M = 88, Md = 91.25, SD = 9.98), exceeding average benchmarks for educational technology tools (typically 60-80)~\cite{vlachogianni2022perceived}. All participants found the 3D environment enhanced their sense of presence and engagement by creating a lecture-like atmosphere that helped maintain focus. Most participants (8/10) valued the note-taking system for capturing ideas during sessions and reviewing knowledge afterward. Following the panel session, some participants (4/10) used the dialogue function to follow up on issues raised during the discussion and noted that the expert agents provided more in-depth clarifications of previously discussed ideas.

The Host Agent was appreciated by all participants for structuring sessions through summaries and segmentation, preventing users from getting lost in complex discussions. However, some participants (4/10) felt it occasionally provided excessive introductory content that reduced overall informativeness. The Audience Agent was highlighted by most participants (7/10) for introducing a student-like, inquiry-driven perspective into the dialogue, prompting experts to propose concrete research methods. As P8 remarked, "\textit{In response to an audience question about how to mitigate cognition dependence, the expert proposed a gamification strategy and elaborated on it by referencing relevant project examples.}"

\section{Discussion}

\subsection{Design Implications of \textit{SimuPanel}}
\subsubsection{Balance between informativeness and human-likeness.}
In our user study, we observed considerable variation in participants’ evaluations of \textit{SimuPanel}'s dialogue content. These differences primarily centered around two aspects: the perceived human-likeness of the simulated dialogue and its level of informativeness. Human-likeness refers to the extent to which \textit{SimuPanel} replicates the dynamics of a real-world panel meeting, including interactional tone among participants~\cite{chaves2021should} and the overall meeting procedure. Informativeness, on the other hand, denotes the degree of concrete detail provided in the dialogues~\cite{fu2023gptscore}, such as thorough definitions and descriptions of referenced projects.

Participants who offered positive feedback commonly appreciated the natural, conversational style adopted by the agents, which they believed enhanced their sense of immersion and identification with the experience. Conversely, some participants expressed a preference for more concise, content-focused dialogues, suggesting that minimizing conversational filler and emphasizing knowledge-rich content could improve learning efficiency.

These findings highlight the importance of balancing human-likeness and informativeness in generative dialogue design, building on Clark’s view that conversational agents should be inspired by human dialogue without merely replicating it~\cite{clark2019makes}. For simulated academic discussions, designers must carefully tune the mix of domain-specific content and conversational flow to support both immersion and effective learning.

\subsubsection{Possible learning scenarios for \textit{SimuPanel}.}
Our study found that participants’ prior knowledge of the discussion topic and their initial learning expectations influenced their overall experience. When users lacked background knowledge about the selected topic, they often reported experiencing cognitive overload and perceived the discussion as too abstract, limiting its educational impact. Such experiences were less common among participants with prior knowledge of the topic. This pattern aligns with established findings in learning science, which indicate that prior knowledge plays a crucial role in shaping learners’ cognitive engagement and overall learning outcomes~\cite{bittermann2023landscape}.

Participants found the \textit{SimuPanel} format particularly valuable for discovering new questions rather than seeking definitive answers. As one participant (P7) noted, “\textit{It helped me think out of the box.}” These participants regarded the simulated discussions as a source of inspiration~\cite{li2024chain, guo2023effects}, offering novel perspectives in their domain of interest. However, those seeking concrete solutions or methodologies often felt that the conceptual nature of the dialogues fell short of their needs.

Taken together, our findings suggest that \textit{SimuPanel} and similar systems are best suited for learners with foundational knowledge who seek to explore new ideas or research directions. They may be less effective for users pursuing concrete problem-solving goals that require structured, methodological guidance.

\subsection{Role-Playing Language Agents Building}

Our ablation study reveals a critical insight for educational agent design: effective expert simulation emerges from integrated cognitive architectures. The substantial performance gap between the Full Reasoning-chain and other approaches demonstrates that expert reasoning components function synergistically, echoing Ericsson's findings in human behavior~\cite{ericsson2018cambridge}, which suggest that expertise arises from integrated knowledge structures, analytical processes, and strategic adaptations.

This synergy extends to persona construction, which we conceptualize not as superficial role-playing but as the epistemic lens through which agents process information. By integrating factual knowledge with epistemological beliefs, \textit{SimuPanel} moves beyond surface-level characteristics toward simulating intellectual identity. Recent research by Hu and Collier~\cite{hu2024quantifying} confirmed this approach, suggesting effective personas must encode not just what agents know but how they evaluate and prioritize information.

The strong performance of our Guided Discourse approach highlights the importance of structured communication strategies in modeling disciplinary discourse patterns. However, a persistent challenge remains in balancing specificity and flexibility—overly generic personas lack differentiation, while rigid ones constrain adaptability across topics~\cite{tseng2024two}.

These findings reframe the conventional treatment of LLM-based agents as mere knowledge repositories. The distinctive value of expert simulation lies in modeling how experts process, evaluate, and communicate knowledge. Future educational systems should prioritize both multi-stage reasoning architectures and modular persona construction that reflect diverse epistemological stances. This approach would not only enhance simulation fidelity but also allow learners to observe how different intellectual paradigms interact, a crucial aspect of authentic academic apprenticeship.

\subsection{Limitations and Future Work}

Our study has some limitations. First, the selected HCI panel topics may not have interested all participants, potentially affecting their engagement and overall evaluation. To make the system more personalized, we could allow users to build expert personas themselves, enabling them to select experts that align with their specific interests and learning goals with improved generalizability. Additionally, this study was conducted using a 3D Unity environment rather than full VR implementation, which could affect user immersion and engagement. Enhanced features such as dynamic camera movement and interactive pop-ups for relevant works like CrossTalk~\cite{xia2023crosstalk} may further improve the learning experience. Lastly, our current approach takes xx seconds to output one utterance, which may limit its real-time application. These approaches would lead to a richer, more personalized user experience in simulated panel discussions and could extend the system's applicability across diverse educational contexts.

\section{Conclusion}

We presented \textit{SimuPanel}, a multimedia learning system that simulates academic panel discussions through LLM-based multi-agent interactions. Grounded in principles of expert representation, structured discourse modeling, and learner engagement, the system is capable of simulating panel discussions centered on specific academic topics. LLM-based evaluations validated the effectiveness of each system component in enhancing the quality of generated dialogue. Findings from the user study further revealed key design factors influencing learning effectiveness and user experience. These insights offer guidance for refining dialogue generation strategies and identifying appropriate application scenarios. The study also highlights the importance of persona modeling and learner agency in educational simulations, pointing toward the development of more personalized and accessible multimedia systems.

\bibliographystyle{ACM-Reference-Format}
\bibliography{sample-base}
\end{document}